\documentclass[12pt]{article} 
\usepackage[sectionbib]{natbib}
\usepackage{array,epsfig,fancyheadings,rotating}
\usepackage[]{hyperref}  

\usepackage{amsmath}
\usepackage{amssymb}
\usepackage{amsfonts}
\usepackage{multirow}
\usepackage{amsthm}
\usepackage{mathrsfs}
\usepackage{bm}
\usepackage{graphicx}
\usepackage{amssymb}
\usepackage{amsmath}
\usepackage{amsmath,amssymb,amsfonts, amsbsy, amsthm, booktabs, setspace,enumerate}
\usepackage{graphicx}
\usepackage[table]{xcolor}
\usepackage{tabularx}
\usepackage[T1]{fontenc}
\usepackage{setspace}
\usepackage{enumerate}
\usepackage{algorithm,algorithmic}
\usepackage[margin=1.0in]{geometry}
\usepackage{authblk}
\usepackage{multirow}

\newcommand{\bfm}[1]{\ensuremath{\mathbf{#1}}}
          
          \def\cB{{\cal  B}}
          
          \def\cD{{\cal  D}}
         
         \def\cF{{\cal  F}}
     \def\bG{\bfm G}     
          
     \def\bI{\bfm I}

          \def\cN{{\cal  N}}

     \def\bT{\bfm T}     
\def\bu{\bfm u}     \def\bU{\bfm U}     
     \def\bV{\bfm V}     
          
\def\bx{\boldsymbol x}     \def\bX{\bfm X}     
\def\by{\boldsymbol y}     \def\bY{\bfm Y}

\newcommand{\bfsym}[1]{\ensuremath{\boldsymbol{#1}}}

\def\bbeta     {\bfsym \beta}
\def\btheta    {\bfsym \theta}
\def\bgamma    {\bfsym \gamma}

\def\bmu       {\bfsym {\mu}}

\def\bomega    {\bfsym \omega}

\def\bGamma  {\bfsym \Gamma}
\def\bTheta  {\bfsym \Theta}

\renewcommand{\hat}{\widehat}

\newcommand{\norm}[1]{\|#1\|}
\renewcommand{\]}{\right]}

\newcommand{\bthetahat}{\hat{\btheta}}

\newcommand{\floor}[1]{\lfloor #1 \rfloor}
\newcommand{\Dscr}{\mathscr{D}}

\newtheorem{theorem}{Theorem}
\newtheorem{lemma}{Lemma}

\theoremstyle{definition}
\newtheorem{definition}{Definition}

\begin{document}


\renewcommand{\baselinestretch}{2}





\fontsize{12}{14pt plus.8pt minus .6pt}\selectfont \vspace{0.8pc}
\centerline{\large\bf Model Selection Confidence Sets by Likelihood Ratio Testing}
\vspace{.4cm} \centerline{Chao Zheng$^1$, Davide Ferrari$^2$ and Yuhong Yang$^3$} \vspace{.4cm} \centerline{\it $^1$ Lancaster University, 
$^2$University of Melbourne and $^3$University of Minnesota} \vspace{1.5cm} \fontsize{10}{11.5pt plus.8pt minus
.6pt}\selectfont


\begin{quotation}
\noindent {\bf Abstract:}
The traditional activity of model selection aims at discovering a single model superior to other candidate models. In the presence of pronounced noise, however, multiple models are often found to explain
the same data equally well. To resolve this model selection ambiguity, we introduce the general approach of model selection confidence sets (MSCSs) based on likelihood ratio testing. A MSCS is defined as a list of models statistically indistinguishable from the true model at a user-specified level of confidence, which extends the familiar  notion of confidence intervals to the model-selection framework.
Our approach guarantees asymptotically correct coverage probability of the true model when both sample size and model dimension increase. We derive  conditions under which the MSCS contains all the relevant information about the true model structure. In addition, we propose natural statistics based on the MSCS to measure importance of variables in a principled way that accounts for the overall model uncertainty. When the space of feasible models is large, MSCS is implemented by an adaptive stochastic search algorithm which samples MSCS models with high probability. The MSCS methodology is illustrated through numerical experiments on synthetic data and real data examples.\par

\vspace{9pt}
\par
\end{quotation}\par

\def\thefigure{\arabic{figure}}
\def\thetable{\arabic{table}}

\renewcommand{\theequation}{\thesection.\arabic{equation}}

\fontsize{12}{14pt plus.8pt minus .6pt}\selectfont

\setcounter{equation}{0} 
\section{Introduction}
Likelihood  inference is a centerpiece of statistical theory and plays an important role in many research fields.  Numerous methods
relying on likelihood objective functions  have been developed in the literature of model selection, ranging from classic information
criteria to more recent sparsity-inducing penalization methods; see \cite{mcquarrie1998regression}, \cite{claeskens2008model} and \cite{buhlmann2011statistics} for book-length expositions.
In the presence of noise in data, however, it is typically difficult to declare a single model significantly superior to all possible
competitors, due to the prevailing effect of the model selection uncertainty. In this situation, multiple or even a large number of models
may be equally supported by data, so that any selection procedure is likely to pick at random a single model from a large set of more or less equivalent models.
Clearly, this implies tossing away valuable information; for example, in regression analysis, alternative combinations of
predictors  may be discarded, whilst such combinations may contain scientifically valid explanations of the phenomenon under examination.

Motivated by the above issues, there has been a growing interest in developing statistical measures of model selection uncertainty. 
The approach followed in this paper proposes to construct a model selection confidence set (MSCS), defined as a set of models  indistinguishable from the true model at a user-defined confidence level. Simply put, the MSCS extends the familiar frequentist notion of confidence intervals to the model-selection framework. \cite{Ferrari&Yang15} first introduced confidence sets for variable selection in the context of linear models by F-testing. They achieve the exact coverage probability for the globally optimal model from the model space. Thus, this is the first work introducing confidence sets in the frequantist sense for variable selection. Moreover, in their framework the number of predictors can grow with the sample size, so that the number of potentially useful models is allowed to be large.

 Different from the MSCS of \cite{Ferrari&Yang15}, \cite{Hansen11} studied another methodology called model confidence sets.  Their approach builds on classic step-down procedures for multiple  hypothesis testing  \citep{ lehmann06, romano05} but starts from a pre-specified user-defined set of models, which has limited sizes.  Step-wise equivalence testing was carried out under a user-defined loss function, followed by an elimination rule to drop  the worst performing models. Previously, \cite{shimodaira98} constructed confidence sets containing models with AIC values near the 
smallest among the candidate models. We refer to the Section 6 of \cite{Ferrari&Yang15} for a detailed discussion.

In this paper, we introduce a general methodology to construct model selection confidence sets via likelihood inference.
We begin by considering a full model with $p$ variables to form a reference model or full model, where $p$ is required to be less than $n$. This preliminary step can be achieved by  any over-consistent model screening method, which selects the relevant variables plus a few other variables. We then test candidate sub-models against the full model, by a likelihood ratio test (LRT) at the significance level $0<\alpha<1$. The MSCS is formed by all the candidate models that survive the LRT screening. This way of construction guarantees
that the globally optimal model is included in the MSCS with probability at least $1-\alpha$ as the sample size increases (under appropriate regularity conditions). From a theoretical viewpoint, we investigate the condition for the MSCS to contain all the relevant information about the model structure when both $p$ and $n$ diverge. Since in practice the MSCS cannot be computed by exhaustive search unless $p$ is very small, when the model space is moderate or large, we propose a  stochastic algorithm (MSCS-AS) which samples MSCS models with high probability.

The proposed MSCS methodology can be used for various tasks in support of the model selection activity. First, given a model selected from some rule, one can immediately use the MSCS to check if such a model is too parsimonious in terms of missing important variables. Second, the frequency of  variables
in the  MSCS can be used to rank their usefulness in a principled way that accounts for the model selection uncertainty.
Third, the MSCS and the associated importance measures may be used to narrow down the list of candidate models by considering the most important variables.

The rest of the paper is organized as follows. In Section \ref{Sec2}, we describe the main MSCS methodology and study the
condition needed to learn the true underlying model structure. In the same section, we propose a measure of importance for the individual variables. In Section \ref{sec:ce_sampling}, we give an
adaptive sampling algorithm that implements the MSCS methodology. In Section \ref{Sec:MC}, we study the finite sample
properties of MSCS by  Monte Carlo simulations for various models. In Section \ref{Sec:realdata}, we  illustrate the MSCS procedure using the
European E.coli outbreak data and the Australian breast cancer family study  data. In Section \ref{Sec:Conclusion}, we conclude and give final remarks.
Technical proofs are deferred to the Appendix.

\setcounter{equation}{0} 
\section{Model selection confidence sets} \label{Sec2}

Consider independent observations, $\bY_1, \dots, \bY_n$, from a family of models indexed by the parameter $\btheta=(\theta_1,\dots,\theta_p)^T\in \bTheta$, with corresponding log-likelihood function $\ell_n(\by,\btheta)$. Each
 parameter element  $\theta_j \in \btheta$ describes a possibly relevant part of the overall model structure. We suppose that only a  subset of
$\btheta$ is useful for describing the data, while the others are regarded as unnecessary. A generic model index $\bgamma$ is defined as a
subset of indexes in $\{1,\dots,p\}$ and we write the correspondent parameter space as
$\bTheta_{\bgamma}$. Denote $\btheta_{\bgamma} \in\bTheta_{\bgamma}$ as a parameter with the model $\bgamma$, and let $p_{\bgamma} = \text{card}(\bgamma)$ denotes
the cardinality (number of elements) of $\bgamma$.
The true parameter vector and the true model is denoted by $\btheta^\ast$ and
$\bgamma^\ast$, repectively, while the full model with $p$ parameters is denoted by
$\bgamma_f$. 
The space of feasible candidate models is  $\Gamma$ which contains the true model $\gamma^\ast$.  
The cardinality of $\Gamma$ may be as large as $2^p$; however it also may be restricted
in some special problems.

In the rest of the paper, we assume
$p<n$, but $p$ is allowed to slowly grow with
$n$, reflecting the notion that with more
observations available, the statistician is tempted to introduce additional variables into the model. For simplicity,
we omit the sub-index $n$  when it is clear from the context.  In what follows, we use  ``$\lesssim$'' to denote that the left hand side is bounded by the right hand side up to some positive constant independent of $n$. We write $a\gtrsim b$ if $b\lesssim a$.

\subsection{Construction by likelihood ratio testing}
\label{section LRTest}

A MSCS is constructed from the known models space, $\Gamma$, and a criterion to assess models in $\Gamma$ empirically. To screen out implausible
models in the context of maximum likelihood estimation, it is natural to use the likelihood ratio test. Given a
candidate model $\bgamma$, we consider testing the null hypothesis
$H_0: \btheta^\ast \in \bTheta_{\bgamma}$ against the
alternative hypothesis $H_1: \btheta^\ast \notin \bTheta_{\bgamma}$. Then  model $\bgamma$
is rejected if
\begin{equation} \label{LR}
\Lambda_{\bgamma} \equiv 2 \left\{  \ell_n(\bthetahat_{\bgamma_f}) - \ell_n (\bthetahat_{\bgamma}) \right\} \ge q(\alpha; p - p_{\bgamma}),
\end{equation}
where:  $\hat{\btheta}_{\bgamma}$ and $\hat{\btheta}_{\bgamma_f}$ denote, respectively, the MLEs for the candidate and full models; $\ell_n(\cdot)$ is the log-likelihood function; $q(\alpha;
d)$ is the upper $\alpha$-quantile for the central chi-squared distribution with $d$ degree of
freedom. The $(1-\alpha)100\%$--MSCS is defined by the set of all models surviving the LRT screening:

\begin{equation}\label{Gammahat}
\widehat{\Gamma}_\alpha \equiv \left\{ \bgamma \in \Gamma:  \Lambda_{\bgamma} \leq q(\alpha; p - p_{\bgamma}) \right\}.
\end{equation}
The LRT  procedure is applied to all models $\bgamma \in \Gamma$: If a model is rejected, then we have evidence that it is
too parsimonious in the sense that it is likely to miss at least one important variable. By default, the
full model  $\bgamma_f$ is included in $\widehat{\Gamma}_\alpha$.

When $p$ is fixed, and $\bgamma^\ast$ is a proper subset of
$\bgamma$,
the limiting null distribution of the LRT statistics $\Lambda_{\bgamma}$
is a central chi-square distribution, which follows directly  from  Wilks
theorem (e.g., see \cite{van2000asymptotic}). By construction, this
implies that the true model is  in the MSCS with
probability approximately $1-\alpha$ in large samples. Specifically, if the true model is not the full model $\left(\bgamma^\ast\neq\bgamma_f\right)$,   we have:
\begin{equation} \label{coverage}
\lim_{n \rightarrow \infty} P( \bgamma ^{\ast }\in \widehat{\Gamma}_\alpha) = 1-\alpha.
\end{equation}
If $\bgamma^\ast=\bgamma_f$, then $P(
\bgamma ^{\ast }\in \widehat{\Gamma }_\alpha ) =1$.

When $p$ increases with $n$, similar Wilks-type results are given by \cite{portnoy1988asymptotic}, \cite{murphy1993testing}, \cite{fan2004nonconcave} and \cite{fan2016guarding}
 for exponential family models, Cox regression, penalized likelihood and goodness of spurious fit for GLMs. These results yield asymptotic coverage probability as in  (\ref{coverage}).   \cite{spokoiny2012parametric,spokoiny2013bernstein} establish Wilks-type behaviours for rather general families of models, which quantify and explicitly describe the error term in the approximation of the likelihood ratio statistics under mild regularity conditions on the parametric family.

 We remark that  although the MSCS includes $\bgamma^\ast$ with at least probability $1-\alpha$, one cannot simply conclude that a variable is important just because it appears in some of the models in  $\widehat{\Gamma}_\alpha$. Actually, unimportant variables tend to appear with a respectable frequency in the MSCS models since larger models containing the true model plus other irrelevant variables is likely to survive the LRT screening. 

\subsection{Asymptotic detectability} \label{Sec:detect}

In this section, we study the conditions under which
the variables in the true model appear with large
frequency in MSCS. The results presented in this section extend the
analysis given by \cite{Ferrari&Yang15} for linear models.

\begin{definition}[Asymptotic  detectability] \label{AD}
The MSCS $(\widehat{\Gamma}_{\alpha})$ is said to asymptotically detect all the true variables, if all the variables in the true model $\bgamma^\ast$ are included in each of the models in $\widehat{\Gamma}_{\alpha}$, with probability going to 1.
\end{definition}

The concept of detectability is closely related to the power of the LRT. In the fixed $p$ scenario when a candidate model
$\bgamma$ misses at least one important variable, under appropriate regularity conditions ensuring asymptotic normality of the MLE, the 
$\Lambda_{\bgamma}$ converges in distribution to a non-central chi-square random variable with degree of
freedom $d_{\bgamma}= p-p_{\bgamma}$. Let $\btheta^\ast_{\bgamma}$ denote the parameter value in the model $\gamma$ that minimizes the Kullback-Leibler divergence from the true density (hence providing the best approximation to the true density).  Then the non-centrality parameter of the asymptotic chi-square distribution is
$
\delta_{\bgamma} = (\btheta_{\bgamma}^\ast-\btheta^\ast)^T
 \mathscr{F}(\btheta^\ast)(\btheta_{\bgamma} ^\ast-\btheta^\ast),
$
where the $\mathscr{F}(\btheta)=-E\left[\partial^2\ell_n(\btheta)/\partial \btheta^2\right]$ is the Fisher information matrix.
In the normal regression case as in \cite{Ferrari&Yang15}, the asymptotic distribution is exact. 
Clearly, when applying LRT to a model $\bgamma$, a large value of $\delta_{\bgamma}$ makes it easier to reject $\bgamma$.

In the following theorem, we show that the limiting non-central chi-square alternative distribution is still valid under certain conditions for an exponential model where $\bY_1, \dots, \bY_n$ are i.i.d. observations from the pdf
\begin{equation}
f(\by;\btheta)= \exp\left[ \btheta^T \by- A(\btheta)\right].\label{pdf.model}
\end{equation}  with respect to a sigma-finite dominating measure. For this model, the true model $\bgamma^\ast$ is defined as the indexes of non-zero component of $\btheta$. Appropriate generalizations of the following result may be derived for other models but they are not pursued in this paper.

\begin{theorem}
Assume
conditions (A1)--(A3) given in the Appendix. Let $\bgamma$ be a model
missing at least one variable in the true model $\bgamma^\ast$ and $d_{\bgamma}\rightarrow \infty$ as $n$ grows. Moverover, assume
$\norm{\btheta^\ast-\btheta^\ast_{\bgamma}}\gtrsim \sqrt{p/n}$ and $p=o(n^{2/3 })$.  Then  for model (\ref {pdf.model}) we have
\begin{align}\label{noncen.appr}
\dfrac{\Lambda_{\bgamma}-(d_{\bgamma}+\delta_{\bgamma})}{ \sqrt{2d_{\bgamma} + 4 \delta_{\bgamma}} } \overset{\Dscr}{\rightarrow}
\cN_1(0,1),\ \ \text{as } n\rightarrow\infty.
\end{align}
\end{theorem}\label{thm.noncen}
Denote $X_{\delta, d}$ as  a chi-square random variable  with degree
of freedom $d$ and non-centrality parameter $\delta$. Recall that $(X_{\delta,d}-d-\delta)/\sqrt{2(d+2\delta)}$ converges to a standard normal distribution $\cN_1(0,1)$, when
$d\rightarrow\infty$. This means that in view  of (\ref{noncen.appr}),  $\Lambda_{\bgamma}$ is approximately non-central chi-square variable with degree of freedom $d_{\bgamma}$ and non-centrality parameter $\delta_{\bgamma}$.

The non-centrality parameter $\delta_{\bgamma}$ may be interpreted a the discrepancy measure due to missing important variables in the true model. From this viewpoint, the relative magnitude of
$\delta_{\bgamma}$ provides us with some insight on how informative is the data in relation to the feasibility of the model selection task.
Let $K_n(s)=s\log(p/s)$. With $p\rightarrow \infty$, it is typically the case that the true model dimension is bounded away significantly from $p$. In the rest of the paper, we assume that $d_\gamma$ increases to $\infty$ (however slowly) uniformly for the candidate models. The following result gives explicit sufficient conditions involving $\delta_{\bgamma}$ for detectability in the general parametric setting, which includes model (\ref{pdf.model}) described in Theorem \ref{thm.noncen}.

\begin{theorem}\label{thm3}
Let $\Gamma_u$ denote the set of models missing at least one of the true variables.
Suppose that it holds that for all $\bgamma\in\Gamma_u$, we have
\begin{equation}\label{A0}
|P\left(\Lambda_{\bgamma}\le q(\alpha;k)\right)-P\left(X_{\delta_{\bgamma}, k}\le q(\alpha;k)\right)|\le c_1\exp\left[-c_2K_n(k)\right],
\end{equation}
where $c_1$ and $c_2$ are positive constants.
A sufficient condition for asymptotic detectability is
\begin{equation}
\min_{\bgamma\in\Gamma_u}\dfrac{\delta_{\bgamma}}{K_n(d_{\bgamma})}>B \label{suffcient},
\end{equation}
for some large enough positive constant $B$.
\end{theorem}

The additional assumption above  requires an exponential probability bound for the chi-square approximation of LRT statistics with model misspecification. We refer to Theorem 3.10 and Proposition B.1 in \cite{spokoiny2013bernstein}, for the chi-square approximation, where a similar bound can be achieved for certain i.i.d and regression models. Moreover, consider normal linear regression as in \cite{Ferrari&Yang15}, $\left(RSS_{\bgamma}-RSS_{\bgamma_f}\right)/d_{\bgamma}$, the numerator of their $F$-test, follows an exactly non-central chi-square distribution where the assumption is trivially satisfied.

\subsection{Sharpness of the sufficient condition for detectability}
 The above detectability condition theorem is a general extension of Theorem 2.3 in \cite{Ferrari&Yang15}, where in the context of normal linear regression, a sufficient condition for detectbility is given as $\min_{\bgamma\in\Gamma_u}\delta_{\bgamma}/\left\{\xi_n+\sqrt{K_n(d_{\bgamma})}\right\}$ is greater than some large enough constant, where $\xi_n\rightarrow \infty$ is any arbitrarily slowly growing sequence. It turns out the condition is in fact not sufficient and a error occurred in their derivation. A correct sufficient condition is that $\min_{\bgamma\in\Gamma_u}\delta_{\bgamma}/K_n(d_{\bgamma})$ is larger than some constant, which matches (\ref{suffcient}) in this paper. In this subsection, we show the new sufficient condition cannot be generally improved. Due to space limitation and the need to correct Theorem 2.3 of \cite{Ferrari&Yang15}, we focus on the normal regression case here. A generalization to other models, e.g. GLMs, can be done similarly with additional technical developments.

Clearly the detectability condition relates to the size of the coefficients. For the following results, we assume the sparse Riesz condition (SRC) \citep{zhang2010nearly} holds and consider $0<\alpha<1/2$.

Let $r^*\leq p/2$ be a positive integer as the number of non-zero coefficients in the true model. Write $f_{\bbeta}(\bX)=\bX^T\bbeta=\sum_{j=1}^p \beta_jX_{j}$ and let $\cB=\left\{\bbeta: \norm{\bbeta}_0=r^* \,\,\text{and}\,\,\norm{f_{\bbeta}}_n^2\le cK_n(r^*)  \right\}$ for some small constant $c>0$, where $\|{\cdot}\|_0$ denotes the $\ell_0$-norm, and $\norm{f_{\bbeta}}_n^2=\sum_{i=1}^n f^2_{\bbeta}(\bX_i)$ with $\bX_i$ being the covariate vector for the $i$-th observation. It represents all linear regression models $\bgamma$ with only $r^*$ non-zero coefficients.

\begin{theorem}\label{thm.sharpness}
Let $\cD$ denote the event that all the variables in $\bgamma\ast$ are included in each of the models in the MSCS. Then when $c$ is small enough, we must have
$$
\limsup_{n\rightarrow\infty} \inf_{\bbeta\in\cB}P_{\bbeta}(\cD)<1.
$$
\end{theorem}


From the theorem, for the true models of dimension $r^*$ with $\norm{f_{\bbeta}}_n^2\le cK_n(r^*)$, detection of the true terms is impossible in a proper minimax sense. Note that, for instance, for the model $\gamma$ that contains all the wrong variables and none of the true variables, it results in 
$d_\gamma =r^*$ and the noncentrality parameter is of order $K_n(d_{\bgamma})$. This matches the lower bound requirement (\ref{suffcient}) in Theorem \ref{thm3} in order. So from this aspect, the sufficient condition (\ref{suffcient}) for detectability cannot be generally weakened in order.

\subsection{Inclusion importance} \label{sec:RIW}

Under the detectability conditions in Theorem \ref{thm3}, the MSCS includes all the relevant information concerning the model selection variability.
Thus, a natural measure for ranking the importance of each parameter element $\theta_k\in \btheta$ is its relative frequency over all
the MSCS models. This suggests the following definition.
\begin{definition}[Inclusion Importance]
The inclusion importance ($II$) for any $\theta_k \in \btheta$ is defined as
\begin{equation}\label{omega_k}
II_k =  \sum_{\bgamma \in \widehat{\Gamma}_\alpha} I(k \in {\bgamma} ) /\rm{card}( \widehat{\Gamma}_\alpha).
\end{equation}
\end{definition}
When $\theta_k$ appears in all MSCS models, its
importance is $II_k=1$, meaning that $\theta_k$ is most likely part of the true model.

As already mentioned, however, we note that a variable cannot be declared  relevant just because it has a non-zero importance index. Actually, unimportant
variable of $\btheta$ tend to appear in the MSCS with frequency near $1/2$. The reason is that when a small model is included in
$\widehat{\Gamma}_\alpha$, also larger models containing the
same variables plus some others tend to be included via the LRT by construction. The following theorem describes an asymptotic
behavior for the inclusion importance.

\begin{theorem}\label{coro1}
If the asymptotic detectability conditions in Theorem \ref{thm3} are satisfied, we have:
\begin{itemize}
\item[\rm {(i)}]$\lim_{n \rightarrow \infty} P(II_k = 1 ) = 1$, for all $k\in \bgamma^\ast$;
\item[\rm{(ii)}]$\lim_{n \rightarrow \infty}P\left(II_k > \dfrac{1}{2}+\Delta\right) \leq \dfrac{\alpha(1+2\Delta)}{4\Delta}$,
 for all $k  \notin \bgamma^\ast$, where $0<\Delta<\dfrac{1}{2}$.
 \end{itemize}
\end{theorem}

If we have sufficient information to learn all the the relevant variables of the true model, we expect that their importance to be close to 1, while the
unimportant variables are not likely exceeding by much the value $0.5$. The upper bound in Theorem \ref{coro1} can be used as a guidance
to control the error probability of over selection. For example, one can set the error probability
$\epsilon = \alpha(1+2\Delta)/(4\Delta)$ to be some small number and then find the corresponding $\Delta$ so as to use II for an understanding if a variable is really important. For example, if the significance level is
$\alpha=0.05$, setting $\Delta=1/6$ implies $\epsilon \leq 0.1$.
\subsection{The  multivariate normal location model}\label{Sec: comment}

In this section  we consider the special case of the multivariate normal distribution with unknown location.
Let $\bY$ follows the $p$-variate normal distribution $\cN_p(\bmu, I_p)$.
Then the pdf with form (\ref{pdf.model}) can be obtained by setting the parameter vector as
$\btheta=\bmu$
and the cumulant generating function is $A(\btheta)= \btheta^T\btheta/2+ p\log(2\pi)/2$.
Assume the true parameter $\btheta^\ast$ is sparse with $p_{\bgamma^\ast}=o(p)$ .
The model space $\Gamma$ is then with cardinality $2^p$.


For a misspecified model $\bgamma\in \bGamma_u$,  the
corresponding non-centrality parameter is $\delta_{\bgamma}=n\sum_{j\in\bgamma^\ast\,\text{and}\, j \notin\gamma}\btheta_j^{\ast2}$. A large value of $\delta_{\bgamma}$  enables us to detect inadequacy of such models.  For example, the asymptotic
detectability conditions in (\ref{suffcient}) states that as long as  the minimum signal is large enough, $\min\{|\theta_i|, \,i\in \bgamma^\ast\}> B\sqrt{p/n}$ for a large enough positive constant $B$, then all the models in MSCS are expected
to contain all the nonzero parameters with probability going to 1 as $n\rightarrow \infty$. Otherwise if the size of some non-zero parameters in $\btheta^\ast$ is too small, then the LRT has not enough power to screen some wrong models out.

\setcounter{equation}{0} 
\section{Implementation by adaptive sampling} \label{sec:ce_sampling}

Testing all the models in $\Gamma$ is computationally challenging unless $p$ is small, since the cardinality of the model
space may grow exponentially in $p$. Thus, in order to find models in the MSCS it seems
natural to turn to sampling methods. Let $\bU=(u_1, \dots, u_p)^T \sim p(\bu;\bomega)$ be a random binary vector representing a model sampled from
$\Gamma$ ($u_j=1$ if the $j$th variable is included in the model), and $p(\cdot;\bomega)$ is a user-defined pmf indexed by $\bomega$. Our main
objective is to choose a value of the parameter $\bomega$  that maximizes the probability to sample MSCS models
\begin{equation}\label{eq:indicator2}
P(\bU \in \widehat{\Gamma}_{\alpha})  = \sum_{\bu \in \Gamma} p(\bu;\bomega)I(\bu \in \widehat{\Gamma}_{\alpha}).
\end{equation}
Note one is unlikely to find models in $\widehat{\Gamma}_{\alpha}$ just by
sampling from some arbitrary pmf $p(\cdot; \bomega)$, unless $\alpha$ is sufficiently small.
Thus, given a target significance level $\alpha = \alpha^\ast$ (e.g. 0.05) we propose to start from some small initial
confidence level, say $\alpha^{(0)}$, and then construct a sequence of significance levels,
$
0 < \alpha^{(0)} \leq  \alpha^{(1)} \le \cdots  \le\alpha^{\ast}
$,
corresponding to sampling distributions $p(\cdot; \bomega^{(0)}), p(\cdot; \bomega^{(1)}), \dots,p(\cdot; \bomega^{(\ast)})$ increasingly concentrated on the target subspace
$\widehat{\Gamma}_{\alpha^\ast}$.

At each step $t\geq 0$ of our algorithm, the parameter $\bomega$ of the sampling distribution is retrieved by the following weighted likelihood approach. We generate $B$
models $\{\bu^{(t)}_b, b=1,\dots,B\}$ from  $p(\cdot; \hat{\bomega}^{(t-1)})$ and then compute
\begin{equation} \label{eq:wlik_emp}
\hat{\bomega}^{(t)} = \underset{\bomega}{\text{argmax}} \ \sum_{b=1}^B I(\bu_b^{(t)} \in \widehat{\Gamma}_{\alpha^{(t-1)}}) p(\bu_b^{(t)}; \bomega ).
\end{equation}
This finds the pmf $p(\cdot; \hat{\bomega}^{(t)})$ closest to the best subset of previously sampled models in terms of their resemblance to
MSCS models. As $t$ increases and $\alpha^{(t)}$ gradually gets closer to $\alpha^\ast$, $p(\cdot;  \hat{\bomega}^{(t)})$ tends to assign larger probability
to models in $\widehat{\Gamma}_\alpha$.

Since this procedure is useful only when the indicator $I(\bu_b^{(t-1)} \in \widehat{\Gamma}_{\alpha^{(t-1)}})=1$ for a
sufficiently large fraction of sampled models, $\bu_1^{(t-1)},\dots,\bu_B^{(t-1)}$, we propose to increase adaptively the significance level as
$
\alpha^{(t)} = \min\{ \text{p-val}^{(t)}_{\floor{(1-\zeta)B}} , \alpha^\ast \}
$, $0< \zeta <1$, where $\text{p-val}_{\floor{(1-\zeta)B}}$ is the empirical $(1-\zeta)$-quantile computed from the distribution of p-values. This ensures that
the event $\{ \bgamma^{(t)}_{b} \in \widehat{\Gamma}_{\alpha^{(t)} }\}$ is not too rare and occurs with probability of approximately $\zeta$. The proposed approach is closely related to cross-entropy (CE) sampling. See
\cite{rubinstein2004cross} for a book-length exposition on this topic and \cite{costa2007convergence} for convergence analysis.
In our practical implementation, we use $p(\bu; \bomega) = \prod_{j=1}^{p} \omega_j^{u_j}
(1- \omega_j)^{1-u_j}$, which gives a closed-form solution to (\ref{eq:wlik_emp}) and leads to a fast algorithm; all our numerical experiments showed
reliable results with relatively fast convergence. Other choices for $p(\bu; \bomega)$ may
enhance the performance of the algorithm, but they are not pursued here. The following steps outline the stochastic procedure for MSCS construction.

\begin{algorithm}[h]
\protect\caption{:\quad{}\bf{MSCS construction by adaptive sampling} \bf{(MSCS-AS)} \label{algMSCS-AS}}
\begin{enumerate}
\item[0.] Initialize $t=0$ (iteration counter) and $\hat{\bomega}^{(0)}$ (parameter vector for pmf $p(\bu,\bomega)$).
\item[1.] Set $t\leftarrow t+1$. Generate $S^{(t)} = \{\bu^{(t)}_1,\dots, \bu^{(t)}_B\}$ from $p(\cdot; \hat{\bomega}^{(t-1)})$, and compute the
sorted p-values, $\text{p-val}^{(t)}_{(1)} \le  \dots \le \text{p-val}^{(t)}_{(B)}$, by the LRT defined in (\ref{LR}).
\item[2.]  Update $\alpha^{(t)} = \min\{ \text{p-val}^{(t)}_{\floor{(1-\zeta)B}} , \alpha^\ast \}$.
\item[3.] Use models sampled in Step 1, maximize the weighted likelihood as in(\ref{eq:wlik_emp}) by computing:
$$
{c}^{(t)}_j = \dfrac{\sum_{b=1}^B I\{ \text{p-val}^{(t)}_{(b)} > \alpha^{(t)},
\hat{\theta}_{j} \in \bu^{(t)}_b\} }{\sum_{b=1}^B   I\{
\text{p-val}^{(t)}_{(b)} >
\alpha^{(t)}\} }, ~~~j=1,\dots, p,
$$where $\{\theta_j \in \bu \}$ denotes the event that the variable
$\theta_j$ appears in model $\bu$.
\item[4.] Update
$ \hat{\omega}_j^{(t)} \leftarrow  \xi c_j^{(t)}  +  (1-\xi) \hat{\omega}_j^{(t-1)}$ for some constant $0 < \xi < 1$.
\item[5.] Repeat Steps 1--4 until $\alpha^{(t-d)}=  \cdots =\alpha^{(t)}= \alpha^{\ast}$, for some $d$ (e.g. $d=10$). The final MSCS is obtained by drawing
$B^{(T)}$ models  from $p(\bu; \hat{\bomega}^{(T)})$, where $T$ denotes the last iteration.
\end{enumerate}
\end{algorithm}

First, note that Step 4 carries out smoothing
at each iteration; if $\xi=1$, the algorithm avoids smoothing. In our simulations, we found that $\xi<1$
performs better than the non-smooth update
with $\xi=1$ since it prevents occurrences of too many zeros and ones in
situations where $p$ is moderate or large. Smoothing avoids local optima where
some model variables  do not have the chance to be
selected, while others are always selected.  In our experience,
the MSCS-AS algorithm is robust to the choice of
$\xi$, with $\xi=0.2$ performing well across all our
numerical examples.

Second, the MSCS-AS algorithm requires setting the initial
weights $\hat{\bomega}^{(0)}$, and the number of models sampled at each
iteration $B$. We found that the procedure is quite robust to the choices of such
parameters. When no prior information on inclusion importance is available, the
initial probabilities $\hat{\bomega}^{(0)}$
can be set as $\hat{\bomega}^{(0)}=(0.5,\cdots,0.5)$. The performance of the
method, however, can be improved by assigning larger weights
to variables that are known to contain more information about the true model.
The number of models $B$ generated in each iteration should be decided based on
affordable computational resources. However, if $B$ is too small this will affect the accuracy of the weighted likelihood criterion (\ref{eq:wlik_emp}). In all our numerical examples we set $B=300$.

Finally, the constant $\zeta$ prevents overly small p-values in the first few
iterations; thus, it ensures a balanced growth of $\alpha^{(t)}$ and guides the sampling
process towards the MSCS models. The parameter $\zeta$ governs the trade-off
between exploration  and exploitation of the
model space $\Gamma$ and it should be
also fixed based on the available computational resources. In our simulations,  $\zeta=0.25$  is found to work well and is compatible with choices of analogous parameters often found in the
CE literature.

\setcounter{equation}{0} 
\section{Monte Carlo experiments} \label{Sec:MC}

\subsection{ MSCS construction by exhaustive search} \label{subsec：mc_ex-search}
In this subsection, we study the finite sample properties of MSCSs constructed
by exhaustive search on the model space $\Gamma$. We generate  samples from the following four models:

\leftmargini=25mm
\begin{enumerate}
\setlength{\itemsep}{1pt}
 \setlength{\parskip}{0pt}
 \setlength{\parsep}{0pt}
\item[\textit{\underline{Model 1}:}] $p$-variate normal with unknown
location, $\bY = (Y_1, \dots, Y_p)^T \sim \cN_p(\btheta, I)$.

\item[\textit{\underline{Model 2}:}] $p$-variate normal $\bY = (Y_1, \dots,
Y_p)^T \sim \cN_p(0, \Sigma)$, with unknown covariance matrix $\Sigma$. Additionally
we assume that $(Y_1,\cdots,Y_p)$ can be partitioned into independent subvectors ($\Sigma_{i,j}=0$ if $Y_i$ and $Y_j$ belong to different subvectors).

\item[ \textit{\underline{Model 3}:}] Logistic regression: $Y|\bx \sim
\text{Bernoulli}(\pi(\bx))$, $\text{Logit}(\pi(x)) = -\bx^T \btheta$, $\btheta \in
\mathbb{R}^p$.

\item[ \textit{\underline{Model 4}:}] Poisson regression: $Y|\bx \sim
\text{Poisson}(\lambda(\bx))$, $\log(\lambda(\bx)) = -\bx^T \btheta$, $\btheta \in
\mathbb{R}^p$.
\end{enumerate}

In Models 3 and 4, the vectors of covariates are sampled from
multivariate normal distribution $\cN_p(0,I)$ at each Monte Carlo run. For all the above models we
consider the following two scenarios describing the relative size of parameters.

\begin{itemize}
	\setlength{\itemsep}{1pt}
 \setlength{\parskip}{0pt}
 \setlength{\parsep}{0pt}
\item[\textit{\underline{Setting 1}}] (constant parameter size):  The first
$p/2$ parameters have the same size and the others are equal to 0. Specifically, $\theta_j =
\psi$, $j = 1, ..., p/2$, and $\theta_j = 0$, $j = p/2+1,...,p$, where the parameters size $\psi$ is set to be $1$  for Models 1 and 3 and $0.2$ for Models 4, respectively.
For Model 2, $(Y_1,\cdots,Y_p)$ is partitioned into $p/2+1$ groups as $\{Y_1,\cdots,Y_{p/2}\},\{Y_{p/2+1}\},\cdots,\{Y_p\}$. We set $\Sigma_{i,j}=1$ if $i=j$; $\Sigma_{i,j}=0.5$,
$1\le i< j\le p/2$, and $\Sigma_{i,j}=0$ otherwise.

\item[\textit{\underline{Setting 2}}] (decreasing parameter size): The first
$p/2$ coefficients have decreasing size and the others are equal to 0.
Specifically, $\theta_j = \psi/j$, $j = 1, ..., p/2$, and $\theta_j = 0$, $j = p/2+1,...,p$,
where $\psi$ is set as $1,2$ and $0.4$ for Models 1, 3 and 4,
respectively. For Model 2, $(Y_1,\cdots,Y_p)$ is partitioned into $p/2+1$ groups as $\{Y_1,\cdots,Y_{p/2}\},\{Y_{p/2+1}\},\cdots,\{Y_p\}$. We set $\Sigma_{i,j}=1$ if $i=j$;
$\Sigma_{i,j}=0.5/|i-j|$ if $1\le i< j\le p/2$, and $\Sigma_{i,j}=0$
otherwise.
\end{itemize}

The above settings are designed to achieve a small signal relative to the noise, so that the
resulting data are affected by model-selection uncertainty. The
model space consists of $2^p$ models for Models 1, 3 and 4. For Model 2, the
model space is equivalent to all the possible
partitions of the set $\{1,2,\cdots,p\}$ and its cardinality can be given by the Bell number $B_p$. In the following simulations, we use
$p=6,8$ corresponding to $B_6=203$ and $B_8=4140$.

\begin{table}[h]
\centering
\small{\begin{tabular}{lccccccccccc}
\hline
\normalsize{}&~&~&\multicolumn{4}{c}{Setting 1}&~&\multicolumn{4}{c}{Setting 2}\\
~&&$n=$&\multicolumn{2}{c}{100}&\multicolumn{2}{c}{250}&~&\multicolumn{2}{c}{100}&\multicolumn{2}{c}{250}\\
\hline
\\
&$\alpha$&& \multicolumn{8}{c}{Model 1}\\
\vspace{0.3cm}
&&$p=$&$8$&$12$&$8$&$12$&~&$8$&$12$&$8$&$12$\\
Coverage (\%)
&0.10&&91.2&90.0&89.8&90.2&&91.2&89.4&91.0&88.6\\
        ~&0.05&&94.4&95.8&94.4&96.2&&94.8&95.4&95.0&94.4\\
        ~&0.01&&98.4&99.8&99.2&98.4&&99.4&98.8&99.0&99.4\\
  \\
Cardinality
  &0.10&&14.5&58.1&14.4&58.3&&22.7&241.3&15.5&101.3 \\
 ~&0.05&&15.3&61.0&15.2&61.2&&27.2&319.0&16.8&126.5\\
 ~&0.01&&15.8&63.5&15.9&63.4&&37.1&497.8&19.4&184.6\\
\\

&&& \multicolumn{8}{c}{Model 2}\\
\vspace{0.3cm}
&&$p=$&$6$&$8$&$6$&$8$&~&$6$&$8$&$6$&$8$\\
Coverage (\%)
         &0.10&&89.2&88.4&89.2&91.0&&89.0&87.6&88.4&88.8\\
        ~&0.05&&94.6&94.8&94.2&94.4&&94.4&93.2&94.4&94.4\\
        ~&0.01&&98.4&97.8&98.8&99.4&&98.4&99.0&98.6&99.2\\
  \\
Cardinality
  &0.10&&13.5&47.1&13.5&47.0&&13.8&57.1&13.4&46.4 \\
 ~&0.05&&14.3&50.7&14.2&47.3&&15.1&72.7&14.2&49.2\\
 ~&0.01&&15.6&57.9&14.8&46.9&&18.8&126.6&14.8&51.5\\
\hline
\end{tabular}}

\caption{Monte Carlo estimates
of MSCS coverage probability and cardinality by exhaustive
search for
Model 1 and Model 2 under varying confidence level $1-\alpha$, sample
size $n$,  and number of
variables $p$. Results  based on 500 MC
runs.}\label{mc_model1&2}
\end{table}

\begin{table}[ht]
\centering
\small{\begin{tabular}{lcccccccccc}

\hline
\normalsize{}&~&\multicolumn{4}{c}{Setting 1}&~&\multicolumn{4}{c}{Setting 2}\\
~&$n=$&\multicolumn{2}{c}{100}&\multicolumn{2}{c}{250}&~&\multicolumn{2}{c}{100}&\multicolumn{2}{c}{250}\\
~&$p=$&8&12&8&12&~&8&12&8&12\\
\hline
\\
&$\alpha$& \multicolumn{8}{c}{Model 3}\\
\\
 Coverage(\%)          &0.10&86.6&80.0&86.6&87.0&&86.6&83.6&87.4&89.2\\
 ~                 &0.05&92.4&89.6&94.6&91.2&&93.0&91.8&93.3&94.6\\
  ~                &0.01&97.6&97.4&99.2&99.0&&96.6&97.6&98.6&98.6\\
  \\
Cardinality    &0.10&17.5&97.3 &14.3&56.8&&35.4&459.5&18.2&195.3	\\
 ~                 &0.05&20.8&147.8&15.2&60.9&&43.6&612.6&21.5&252.4	\\
  ~                &0.01&32.4&257.0&15.9&67.0&&62.0&941.4&29.2&383.6\\
\\

&& \multicolumn{8}{c}{Model 4}\\

\\
  Coverage(\%)       &0.10&89.0&90.6&90.4&89.8&&90.6&89.2&89.0&90.4\\
 ~               &0.05&95.6&94.2&95.2&94.6&&95.2&94.2&94.6&95.0\\
  ~              &0.01&99.4&99.4&99.0&99.0&&99.8&99.4&98.6&99.0\\
\\
Cardinality  &0.10&85.2 &871.6 &24.6&148.7&&75.6 &1269.3&47.9&597.8 \\
                ~&0.05&109.6&1217.8&32.4&210.3&&94.1 &1633.6&51.0&796.7\\
                ~&0.01&157.9&1987.0&54.8&402.1&&124.9&2177.4&71.3&1167.7\\
\hline
\end{tabular}}

\caption{ Monte Carlo estimates of
MSCS coverage probability and cardinality by exhaustive search for
Logistic regression (Model 3) and Poisson regression (Model 4)
under varying confidence level $1-\alpha$, sample size $n$, and number of
variables $p$.  Results based on
500 MC runs.}\label{mc_model3&4}
\end{table}

Tables \ref{mc_model1&2} and \ref{mc_model3&4} show Monte Carlo estimates for the
coverage probability and cardinality of MSCS corresponding to different sample sizes,
$n$,  number of predictors,  $p$,  at the $90$,
$95$ and $99\%$ confidence levels. As one expects, the cardinality of the MSCS grows
as $\alpha$ decreases, while it increases rapidly with $p$, especially when the sample size $n$
is relatively small. This reflects the situation where the  data contain too much noise
and the subsequent model selection variability is pronounced. The cardinality of the MSCS drops quickly as $n$ increases.

In most cases, the true coverage probability is
quite close to the nominal confidence level. And will be improved in general as the sample szies increases.  We note that the true coverage probability tends to be more off from the nominal level
 when the size of the true
parameters is decreasing (Setting 2).
Clearly, in such settings model selection is more challenging, which leads to a increased cardinality of the MSCS but maintains the same coverage probability.

\subsection{MSCS construction by stochastic search}
In this subsection, we study the performance of the MSCS-AS algorithm described in Section
\ref{sec:ce_sampling}. We generate data from Models 3 and 4 (Poisson and Logistic regression models) using a
setting similar to that in \cite{fan2011nonconcave}. For both models, we set $\btheta=\left(\btheta_1^{T},0,0,\dots,0\right)^T$, where $\btheta_1 =
(2.5,-1.9,2.8,-2.2,3)^T$ in Model 3 and $\btheta_1 = (1.25,-0.95,0.9,-1.1,0.6)^T$ in Model 4. The vector of covariates are
sampled from a multivariate normal distribution $\cN_p(0,\Sigma)$ at each Monte Carlo run, where $\Sigma$ has elements $\Sigma_{i,j}=0.5^{|i-j|}$, $i,j=1,2,\dots,p$. For illustration of capability to handle large $p$, we show the results for $(n,p)$ equal to  $(200,100)$ and
$(1000,500)$. The initial weights for the MSCS-AS algorithm are $\hat{\bomega}^{(0)}=(0.5,\cdots,0.5)$, corresponding to
lack of prior information about predictors' importance. The remaining tuning parameters are set as $\zeta=0.25$, $\alpha^{*}=0.05$, and $\xi=0.2$.
\begin{figure}[ht]
\centering
\label{Power-compare}
\begin{tabular}{cc}
~~~~~~Binomial regression ($p=100$) & ~~~~~~Poisson regression ($p=100$)\\
\includegraphics[scale=0.5]{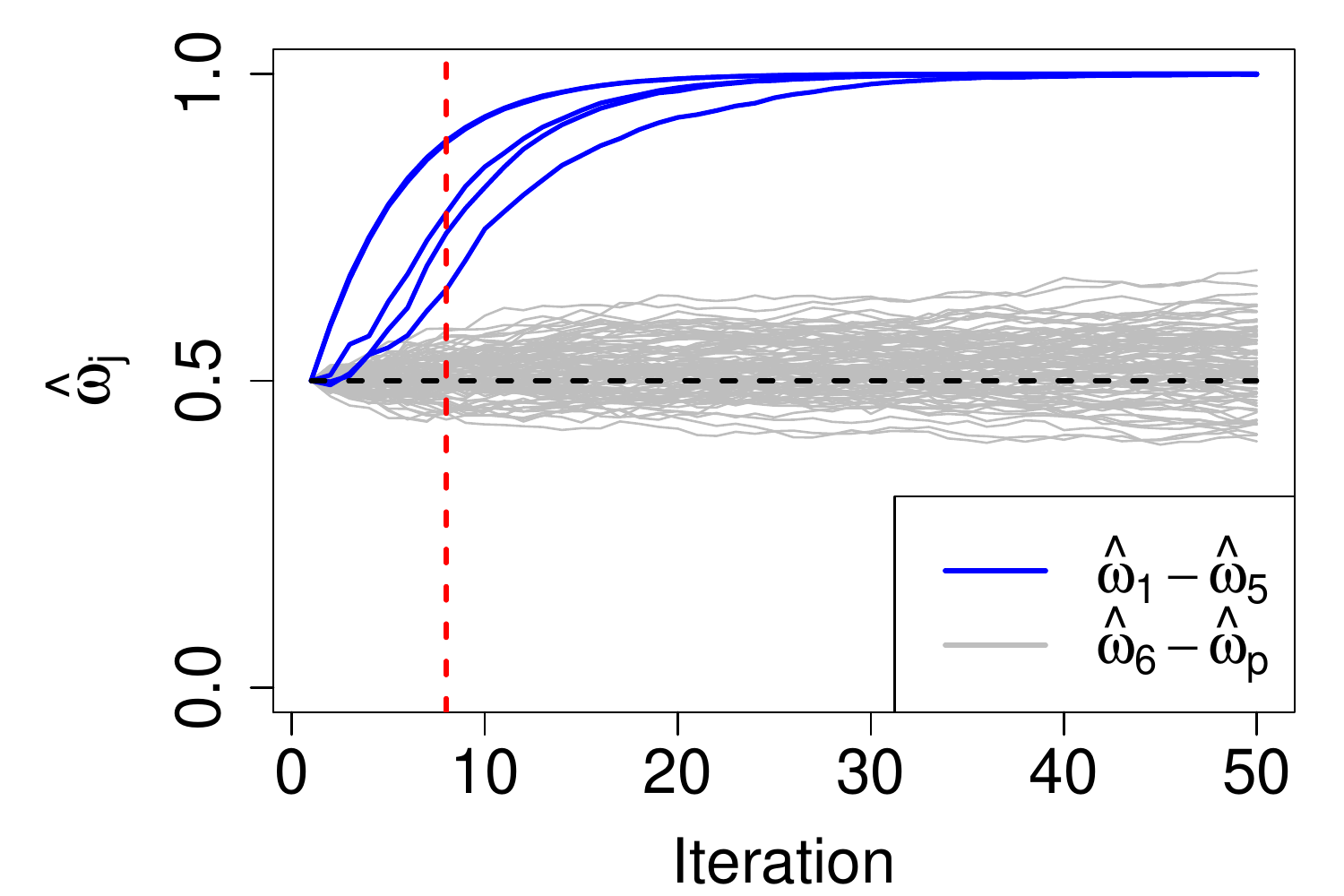}
&\includegraphics[scale=0.5]{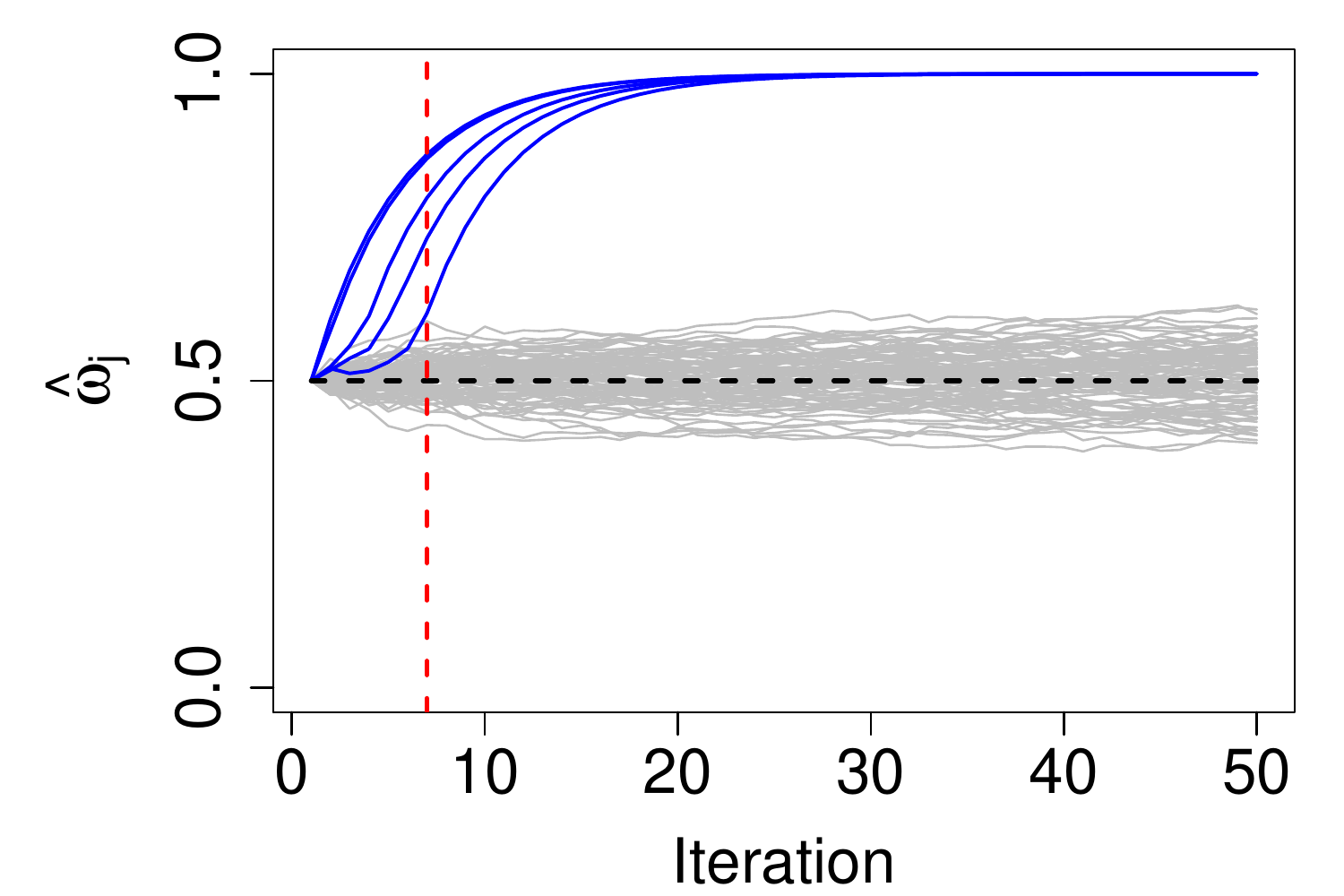}\\
~~~~~~~~Binomial regression ($p=500$) & ~~~~~~~Poisson regression ($p=500$)\\
\includegraphics[scale=0.5]{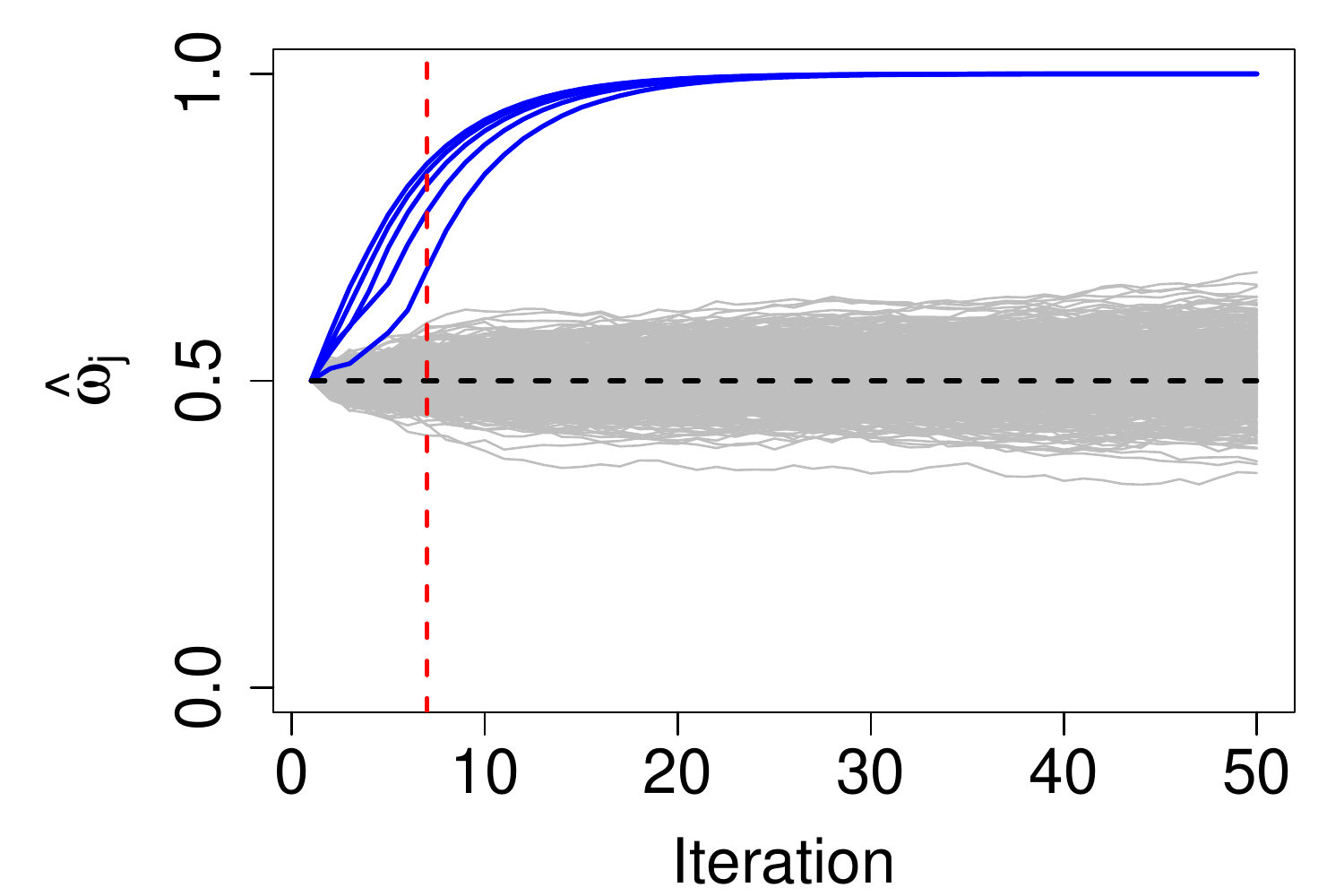}
&\includegraphics[scale=0.5]{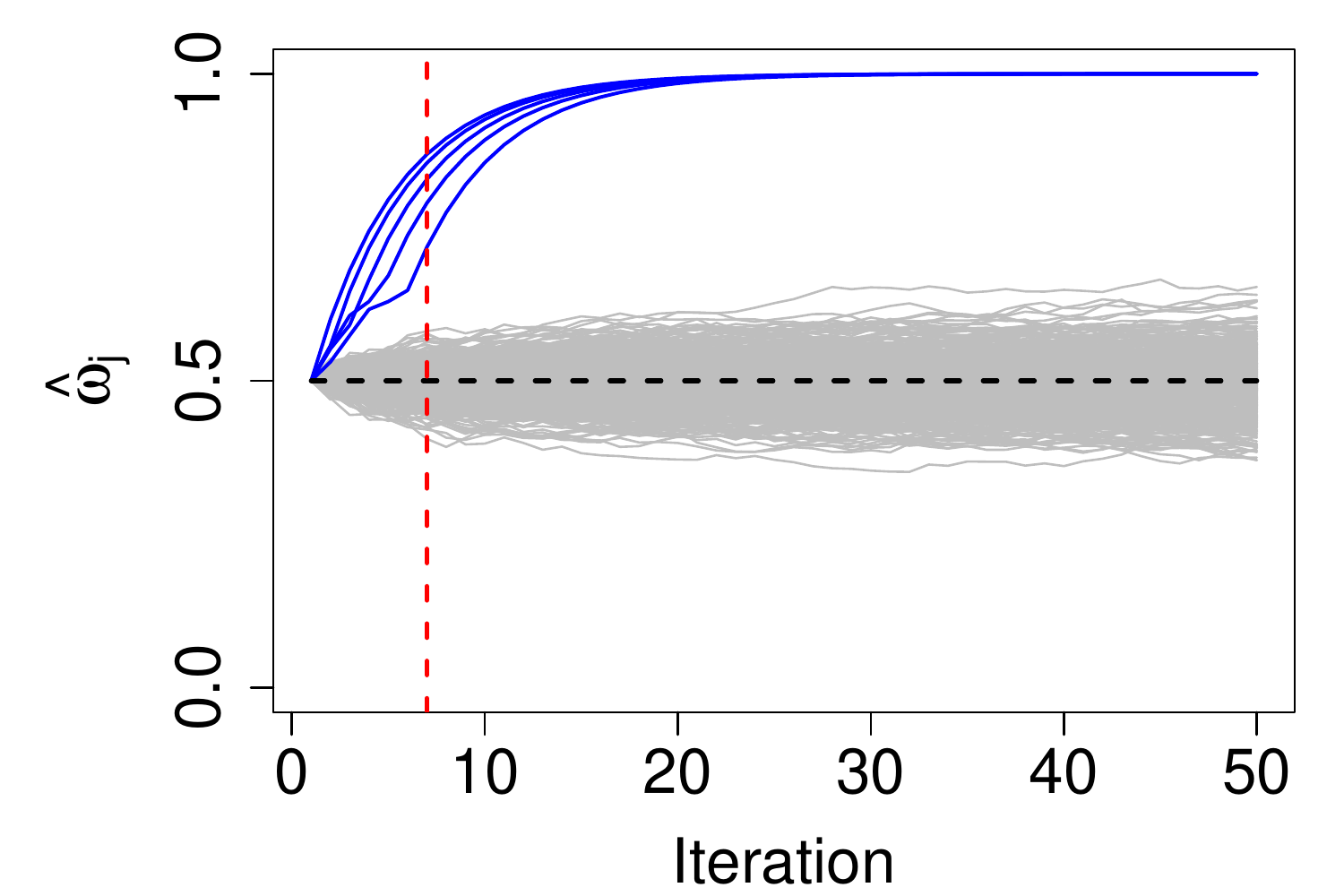}\\
\end{tabular}
\caption{Sampling weights in $p(\cdot; \bomega)$ for 50 iterations of the MSCS-AS algorithm at the 95\% confidence level. The vertical dashed line
corresponds to iteration $t$ such that $\alpha^{(t)}=\alpha^\ast$. Left and right panels correspond to Binomial and Poisson regression models described in Section
\ref{Sec:MC}. Settings for the algorithm: $B=300$, $\zeta=0.25$,
$\alpha^{*}=0.05$, and $\xi=0.2$.}\label{fig1}
\end{figure}

Figure \ref{fig1} shows the trajectories for the importance weights $\{\omega_j\}_{j=1}^p$ during the first $50$ iterations of the algorithm at the $95\%$ confidence level.  In all the considered cases,
the trajectories corresponding to terms with non-zero coefficients are clearly distinguished from the others after a few iterations. The inclusion importance for the relevant terms increases to around $1$, while the
others only have importance weights near $0.5$. The graphs show that the MSCS-AS algorithm samples with probability  progressively concentrating on
the true model terms, while unimportant terms are sampled quite randomly. This behaviour mimics the structure of the true MSCS, $\widehat{\Gamma}_\alpha$, thus enabling us to  detect MSCS models at a much cheaper computational cost than exhaustive
search on $\Gamma$.

As suggested in Section \ref{sec:ce_sampling} (Step 5 of the algorithm), we stop updating the importance weights shortly after the sequence of significance levels
$\alpha^{(0)}, 																\alpha^{(1)}, \dots $ reaches the target significance level $\alpha^\ast$. However, after $\alpha^{(t)}$ reaches $\alpha^\ast$, the weights of the irrelevant predictors eventually  converge to 0 or 1 according to Kolmogorov's zero-one law, thus one should stop before
that happens. For example, stopping shortly after $\alpha^{(t)}=\alpha^\ast=0.05$ -- say around 15 iterations -- already enables us to detect useful predictors from the rest.

To illustrate that the MSCS-AS algorithm generates MSCS models with large probability, we stop at iteration 15 and sample $10^6$ models using $p(\cdot; \hat{\bomega}^{(15)})$, where $\hat{\bomega}^{(15)}$ is the importance weight at the $15$th iteration. For the binomial regression model with $p=100$ predictors, $80.5\%$ of  the models generated are included in MSCS. For the Poisson regression model with $p=100$ predictors, $76.6\%$ of the total models generated are in the MSCS. In comparison, if we generate $10^6$ models using the uninformative weights $\bomega^{(T)}=\left(0.5,\cdots,0.5\right)$, the proportion of MSCS models is basically $0$, due to the largeness of the model space.

\setcounter{equation}{0} 
\section{Real data examples}\label{Sec:realdata}

\paragraph{Example 1: European Escherichia coli(E.coli) O104:H4 outbreak data.} In this example,
we apply the MSCS methodology to the E.coli data as described in \cite{david2016reddog}. E.coli O104:H4 is a particularly
aggressive pathogen and caused a serious outbreak in northern Germany in 2011 \citep{rasko2011origins}. Both during and after the outbreak,
scientists have examined the genome of E.coli to find genetic causes for the severity of the outbreak. The data set used here consists of 56 outbreak isolates.
For each isolate, 10 genes (or hypothetical genes) in the O104:H4 pangenome (a full collection of genes in a species of bacteria) that have been identified as might have been associated with the
outbreak are considered. The main goal of our analysis is to select a model which can explain the most meaningful interaction effects between those genes.

The presence of genes in E.coli is denoted by binary variables taking values 1 when the correspondent gene is present and 0
otherwise.  Let
$\bY=(Y_1,\dots, Y_{10})$ be a random variable with 10 binary variables each denoting the activity of a particular gene. The pmf of $\bY$ is modelled by the
Ising Model,

\begin{equation}
P(\bY=\by;\btheta)=\exp\left(\sum_{1\le j\le k\le 10}\theta_{j,k}y_j y_k+\psi(\btheta)\right), \label{Ising.eq}
\end{equation}
where $\btheta=\left(\theta_{j,k}\right)_{1 \le j\le k\le 10}$ is the parameter of interest with $p=55$ and $\psi(\btheta)$ is the normalizing constant.
The variable $\theta_{j,j}$ is regarded as the main effect for gene $j$, whilst $\theta_{j,k}$ is interpreted as an interaction effect between genes
$j$ and $k$.
Here, we wish to choose $\theta_{j,k}\neq 0$ if genes $j$ and $k$ have interaction (in the same group) and $\theta_{j,k}= 0$ otherwise. In our analysis,
we assume $\theta_{j,j} \neq 0$ for $1\le j\le 10$, meaning that the main effects are always included. The total number of possible models in $\Gamma$ is
$B_{10}=115975$, corresponding to the $10$-th Bell number, which counts the number of different ways to partition a set containing $10$ elements.

\begin{figure}[ht]
\centering
\begin{tabular}{c}
\includegraphics[scale=0.6]{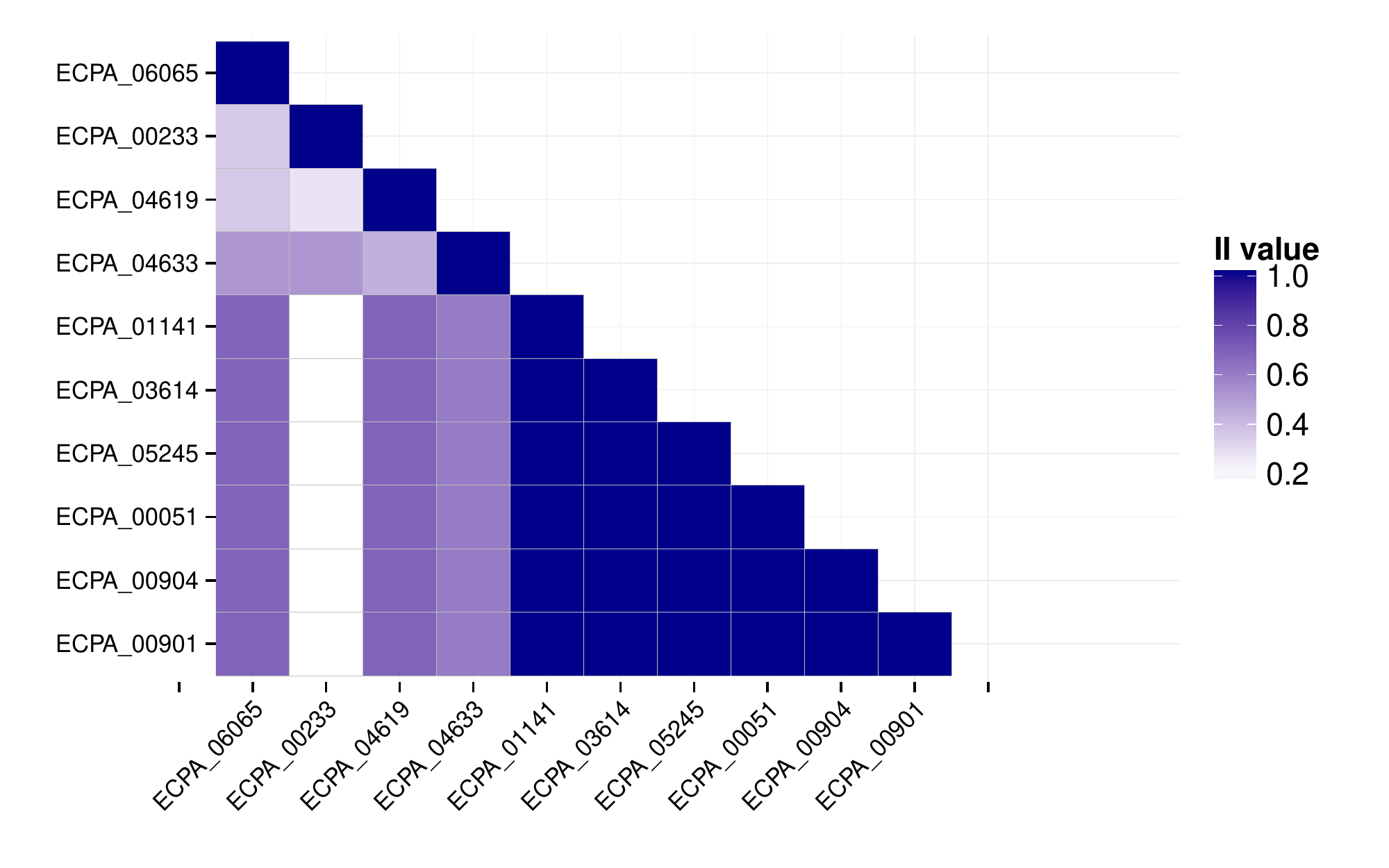}\\
\end{tabular}
\caption{Inclusion Importance ($II$) for $\theta_{j,k}$ in model (\ref{Ising.eq}). The $II$ values are calculated from the $95\%$-MSCS.
}  \label{real_fig1}
\end{figure}

Here we use the exhaustive search to construct the MSCS for $\btheta$. The MSCSs at the $90$, $95$ and $99\%$ confidence levels contain
$7$, $12$ and $38$
models, respectively;   these numbers are   small compared to the model space size, meaning that most of the models in $\Gamma$ are rejected by the LRT
procedure and there is not too much model selection uncertainty here. The $jk$-th element of the matrix in Figure \ref{real_fig1} represents the inclusion
importance ($II$) for the variable $\theta_{j,k}$. Note that certain gene pairs have  high $II$ values; for example, pairwise interactions among
genes 1141, 3614, 5245, 0051, 0904, 0901 are close to $1$  in terms of $II$ values, suggesting that  such genes form a synergetic network associated with
the outbreak occurrence. Other genes, such as 6065, 0233, 4619, 4633, show inclusion importance for interaction effects close or smaller than $0.5$, which
suggests that the corresponding interactions are small or irrelevant.

\begin{figure}[!htbp]
\centering
\begin{tabular}{c}
\includegraphics[scale=0.45]{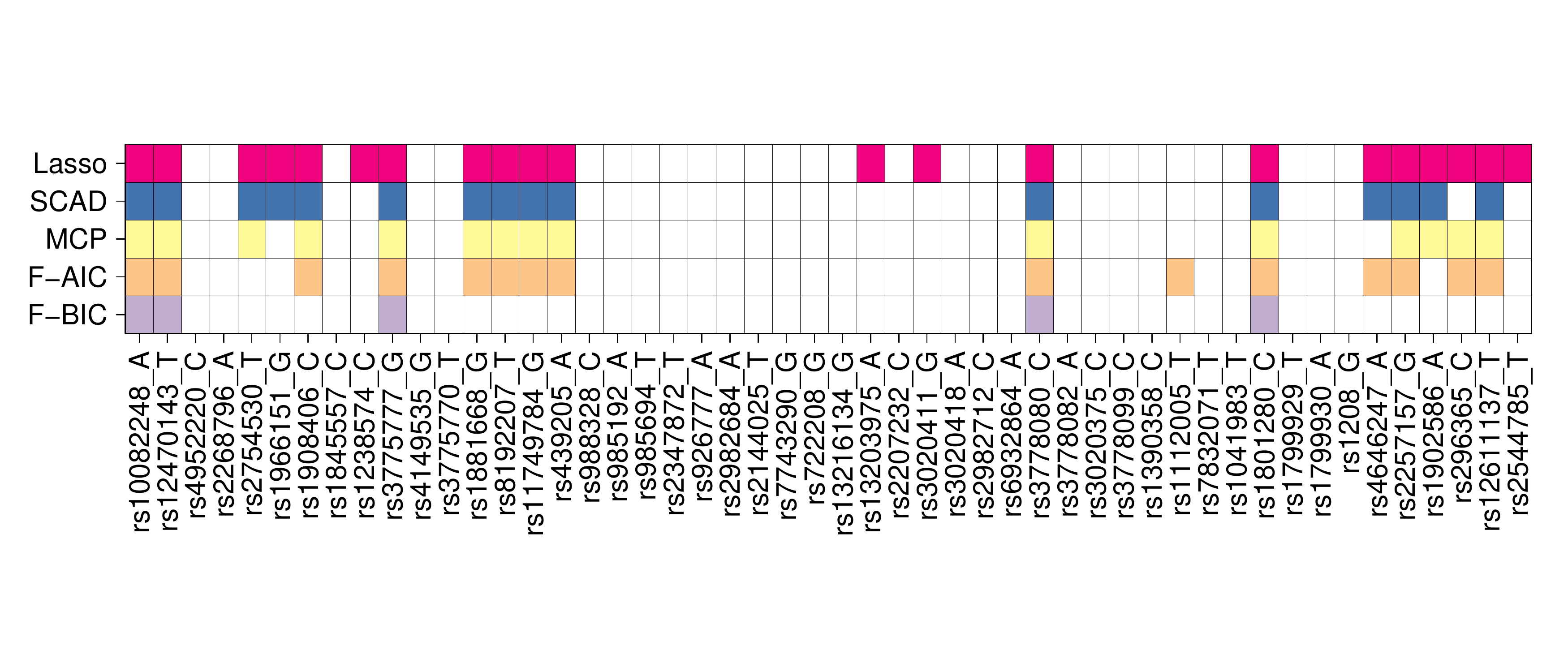} \vspace{-1.5cm}\\
\includegraphics[scale=0.45]{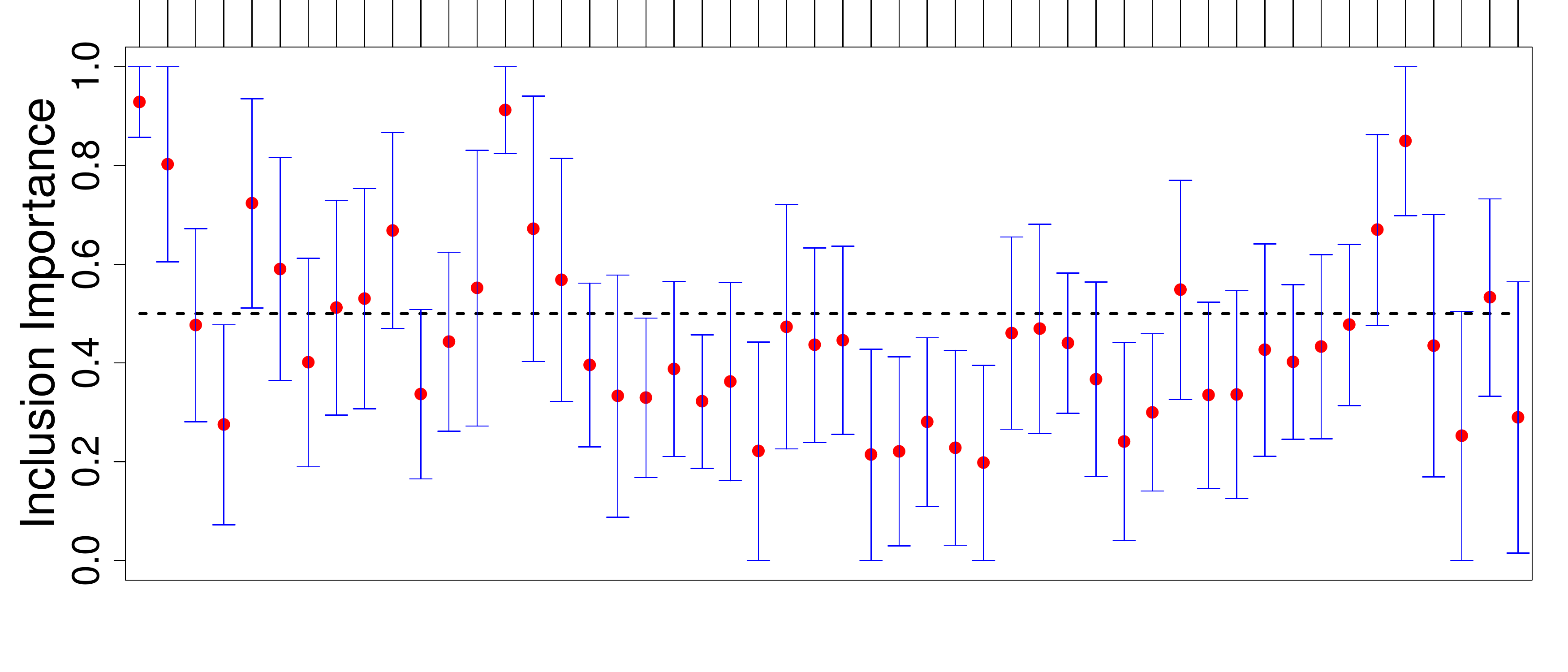}\\

\end{tabular}
\caption{Analysis of the ABCFS case-control genotype data.
Top: model selection by penalized likelihood methods with Lasso, SCAD and MCP penalties and step-wise forward AIC and forward BIC. Colored cells denote selected variables.
Bottom: Estimate of $II$ by MSCS-AS algorithm(red points) and the 95\%
bootstrap confidence intervals. The number of bootstrap replicates is 50.  Settings for the MSCS-AS algorithm parameters: $B=300$,
$B^{(T)}=10^6$, $\alpha^\ast=0.05, \zeta=0.25, \xi=0.2$.}.  \label{real.marginal}
\end{figure}

\paragraph{Example 2: Australian breast cancer family study data.}
 In the second example, we apply the MSCS methodology to the ABCFS genotype data, consisting of $356$ observations ($284$ breast cancer patients and $72$  controls).  Cases are obtained from the Australian Breast Cancer Family Study (ABCFS)  \citep{Dite&al03}, while controls are from the Australian
Mammographic Density Twins and Sisters Study by \cite{Odefrey10}. Patients are genotyped using
a Human610-Quad beadchip array. The response is the binary disease status (presence/absence of breast cancer), while the predictors are 50
SNPs, measured at different loci encoding a candidate susceptibility pathway (probe IDs are listed in
Figure \ref{real.marginal}). To model the binary disease status, we use a logistic regression model. MSCS models are sampled using the MSCS-AS algorithm
described in Section \ref{sec:ce_sampling} with tuning parameters
$\alpha^\ast=0.05$, $\zeta=0.25$, $B=300$ and $B^{(T)}=10^6$.

Figure \ref{real.marginal} (top) shows the models selected by forward step-wise AIC and BIC (F-AIC and F-BIC), and penalized likelihood methods
under Lasso  \citep{tibshirani1996regression}, SCAD \citep{fan2001variable} and MCP \citep{zhang2010nearly} penalties where tuning parameters are all chosen by five-fold cross-validation. Figure \ref{real.marginal} (bottom) shows Inclusion Importance, $II$ values, for each SNP.
The vertical bars represent the 95\% bootstrap confidence intervals for $II$ values. First note
that SNPs with high $II$ values show considerable overlap with those selected by the other methods. Particularly, most
of the SNPs reported as important by more than one model selection method  have large $II$ values. Predictors with $II$ values that are not significantly larger than $0.5$ may not necessarily be important, and need to be consider more carefully.

The p-values of the F-test for the AIC, BIC, Lasso, SCAD and MCP models are $0.96$, $0.09$, $0.88$, $0.86$ and $0.80$.
respectively. Therefore, while the BIC model is included in the MSCS at the $95\%$ and $99\%$ levels, it is not accepted at the $90\%$ confidence level. This sugggests that F-BIC model any misses some important SNP predictors and is not as reliable as other models. Moreover, note that there are only 5 SNPs (rs0082248\_A, rs12470143\_T, rs2754530\_T, rs8192207\_T, rs2257157\_G) that have relatively large $II$ values and confidence interval significantly above $0.5$. Many predictors chosen by some of the methods have $II$ confidence intervals covering $0.5$. This suggests that the sample is not sufficiently informative to
declare such terms relevant so they should be further studied with particular care. To confirm this, we investigate the marginal significance of all the selected SNPs in each of those models. At the $0.05$ level, we have $8$ (Lasso, SCAD and MCP ) to $12$ (AIC) significant coefficients, while the $5$ SNPs with $II$ interval not including $0.5$ are significant in all cases.

\begin{table}[!htbp]
\centering
\begin{tabular}{cccccc} \toprule
 & Lasso & SCAD & MCP &F-AIC &F-BIC\\
\midrule
\midrule
Size  &21 &16  &15  &15   &5      \\
AHD &14.74 & 11.132  & 10.44   &11.58   &5.59 \\
\bottomrule
\end{tabular}
\caption{The number of SNPs in the original selected model and the
average Hamming distance between the $500$ bootstrapped models and the original selected model. }
\label{table.AHD}
\end{table}
Finally, we show the instability of common model selection methods for this dataset. We consider the selected models as in Figure \ref{real.marginal} (top). For each model we obtain the fitted values $\{\hat{Y}_i\}_{i=1}^n$, which are equal to $\{\hat{p}\}_{i=1}^n$, the estimated probability in the logistic regression. Next, parametric bootstrap is used to generate bootstrap replicates $\{X_i, Y_i^\ast\}_{i=1}^n$, and all the methods are applied again to corresponding bootstrap samples. We repeat this step for $S=500$ times and compute the average hamming distances (AHD) of the $S$ bootstrapped Lasso, SCAD, MCP, F-AIC and F-BIC models to the the respective models obtained from the original sample, see Table \ref{table.AHD}. The AHD is large compared the size of the models, which means for each bootstrapped sample, those model selection methods will choose quite different predictors.

%
%

\section{Conclusion and final remarks}\label{Sec:Conclusion}

The MSCS methodology in this paper introduces new tools supporting the activity of model selection in the context of
likelihood-based inference. Since the MSCS is asymptotically guaranteed to contain the true model at a pre-specified confidence level, it represents a natural extension of the familiar notion of confidence intervals to the model selection framework. Furthermore, Theorem \ref{coro1} suggests that important variables
 tend to appear in the MSCS models with large probability as $n\rightarrow \infty$, while unimportant terms appear randomly with frequency not significantly larger than
$0.5$. By looking at the variables appearing frequently in the MSCS one can also choose a single central model representing the entire MSCS by taking predictors with inclusion importance significantly larger
than $0.5$. In the future, developing a theoretical understanding of the optimal way to combine MSCS models would be very valuable as it can
potentially lead to improved  model combining and model selection strategies.

The main focus of the current MSCS approach is based on maximum likelihood estimation. We have shown that in
exponential family models this requires $p=o(n^{2/3})$  and correct model specification for the MSCS to be meaningful. In the future, however, higher-dimensional problems may be pursued, by replacing the LRT statistics with other tools to construct the MSCS, e.g., using penalized likelihood methods. Computational methods to tackle the case where the model space and MSCS is large is also of great interest.

\setcounter{equation}{0} 
\section*{Appendix: Proofs}

Notice that quantities such as $p$, $\btheta$, $\bgamma$ and $\bTheta$ may depend on $n$, hence array asymptotics are considered in this section.

\paragraph{Proof of Theorem \ref{thm.noncen}.} 
Since within the considered exponential family  affine mappings are preserved, without loss of generality we assume $E(\bY_i)=A'(\btheta^\ast)=0$ and $\text{cov}(\bY_i)=A''(\btheta^\ast)=I_{p}$.

Denote $\bV_{\btheta}=\bY_{\btheta}-E(\bY_{\btheta})$, where $\bY_{\btheta}\sim f(\by;\btheta)$. First, we need to assume following regularity conditions for the exponential family model $f(;\btheta)$:
\leftmargini=13mm
\leftmarginii=13mm
\begin{itemize}
	\setlength{\itemsep}{1pt}
 \setlength{\parskip}{0pt}
 \setlength{\parsep}{0pt}
\item[(A1)]   $EY_{ij}^6< \infty, \quad (j=1,\dots,p)$; \label{A1} 
\item[(A2)]$\sup\limits_{\norm{a}=1 \atop \norm{\btheta-\btheta^\ast}^2\lesssim
p/n}\left|E(a^T \bV_{\btheta})^3\right|\lesssim \sqrt{n/p}$,  and $\sup\limits_{\norm{a}=1} \left|E(a^T \bV_{\btheta})^3\right|=O(1)$; \label{A2}
\item[(A3)]  $\sup\limits_{\norm{a}=1 \atop \norm{\btheta-\btheta^\ast}^2\lesssim p/n}E(a^T \bV_{\btheta})^4 =O(1)$.\label{A3}
\end{itemize}

 Suppose we have  $\bgamma$ which is a model not containing all the elements in $\bgamma^\ast$. Let $\overline{\bY}=\sum_{i=1}^n \bY_i/n$ and $\overline{\bY}_{\bgamma}=\sum_{i=1}^n \bY_{i,\bgamma}/n$, where $\bY_{i,\bgamma}$ denote the vector with elements equal to $\bY_{i}$ for indexes in $\bgamma$ and zero otherwise.

Let $\bG=\btheta^\ast-\btheta_{\bgamma}^\ast$  satisfying $ \norm{\bG}\gtrsim \sqrt{p/n}$ and assume $p=o(n^{2/3})$, where $\btheta^\ast_{\bgamma}$ is the $p_{\bgamma^\ast}$-variate vector with components equal to $\btheta^\ast$ at $\bgamma\cap\bgamma^\ast$ and equal to zero otherwise. 
.  To proceed, we give following auxiliary lemmas .

\begin{lemma}\label{lem1}

Suppose conditions (A2) and (A3) hold. Then, for the MLEs, $\bthetahat_{\bgamma}$, we have:
$
\norm{\bthetahat_{\bgamma}-\btheta_{\bgamma}^\ast}=O_p\left(\sqrt{p_{\bgamma}/n}\right)
$,
and
$
\norm{\bthetahat_{\bgamma}-\btheta_{\bgamma}^\ast-\overline{\bY}_{\bgamma}}=O_p\left(p_{\bgamma}/n\right).
$

\end{lemma}

Lemma \ref{lem1} combines Theorems 2.1 and 3.1 in \cite{portnoy1988asymptotic}. 
The existence of MLE for $\btheta_{\bgamma}$ in a $L$-2 neighbourhood of order $\sqrt{p_{\bgamma}/n}$ is still valid by simply changing the true model $\btheta^\ast$ with the partial model $\btheta_{\bgamma}^\ast$ .

\begin{lemma}\label{lem2}
Assume (A2) and (A3) hold,  for model $\bgamma$, $\left|A(\btheta_{\bgamma}^\ast)-A(\btheta^\ast)\right|=\norm{\btheta_{\bgamma}^\ast-\btheta^\ast}^2/2+o(\norm{\btheta_{\bgamma}^\ast-\btheta^\ast}^2)$.
\end{lemma}

\begin{proof}
Note that from (A3), by Taylor expansion we have:
$$
\big|A(\btheta^\ast_{\bgamma})-A(\btheta^\ast_{n})\big|=\Big|\dfrac{1}{2}\norm{\btheta_{\bgamma}^\ast-\btheta^\ast}^2+\dfrac{1}{6}E\left[(\btheta^\ast_{\bgamma}-\btheta^\ast_{n})^T\bV_{\btheta^\ast}\]^3\Big|+O(\norm{\btheta^\ast_{\bgamma}-\btheta^\ast_{n}}^4).
$$
Applying (A2) gives $E\left[(\btheta^\ast_{\bgamma}-\btheta^\ast)^T\bV_{\btheta^\ast}\]^3\lesssim \norm{\btheta_{\bgamma}^\ast-\btheta^\ast}^3$, which completes the proof.
\end{proof}

\begin{lemma}\label{lem3}
Suppose conditions (A1), (A2)and (A3) hold, when $p-p_{\bgamma}\rightarrow\infty$
we have
\begin{eqnarray}
\frac{n\left(\norm{\overline{\bY}}^2-\norm{\overline{\bY_{\bgamma}}}^2 \right)+2n\bG^T\overline{\bY}-(p-p_{\bgamma}  )}{\sqrt{2(p-p_{\bgamma}+2\delta_n)}}\overset{\Dscr}\rightarrow \cN_1(0,1),\label{mar.clt}
\end{eqnarray}
where  $\delta_n=n\norm{\bG}^2$.
\end{lemma}

\begin{proof}

 Define $\bT_k=\sum_{i=1}^{k}\bY_i$, $\bT_{k,\bgamma}=\sum_{i=1}^{k}\bY_{i,\bgamma}$  and $S_k=\norm{\bT_k}^2-\norm{\bT_{k,\bgamma}}^2+2k\bG^T \bT_k-k(p-p_{\bgamma})$. It is easy to see that $E(S_n)=0$ and $\text{var}(S_n)=2n^2(p-p_{\bgamma}+2n\norm{\bG}^2)$.

Let $D_k=S_k-S_{k-1}$, thus
\begin{align}
D_k=2\bY_k^T \bT_{k-1}+\norm{\bY_k}^2-2\bY_{k,\bgamma}^T \bT_{\bgamma,k-1}-\norm{\bY_{k,\bgamma}}^2+2\bG^T \bT_{k-1}+2k\bG^T \bY_{k}-(p-p_{\bgamma}). \nonumber
\end{align}
Next, define $\sigma_k^2=ED_k^2$ and $s_k^2=\sum_{i=1}^k\sigma_i^2$. Note that $\bY_k$ and $\bT_{k-1}$ are independent.
A simple calculation shows that:
 \begin{equation}\label{s_n}
 s_n^2=\sum_{k=1}^n\sigma_k^2\lesssim n^2 p+n^3\norm{G}^2,
 \end{equation}

Next, let $\cF_k=\cF(\bY_1,\dots,\bY_k)$ denote the $\sigma$-field generated by $\bY_1,\dots,\bY_k$. Then $\{S_k\}$ are martingales on $\{\cF_k\}$, and $\{D_k\}$ are the martingale differences. From \cite{chow2012probability}, by Martingale Central Limit Theorem , we have $S_n/\left(n\sqrt{p-p_{\bgamma}+\delta_{\bgamma}}\right)\overset{\Dscr}{\rightarrow} \cN_1(0,1)$ if
\begin{eqnarray}
\sum_{k=1}^nE|D_k|^3/s_n^3&\rightarrow 0 \label{mclt.c1},\quad\text{and}\quad\sum_{k=1}^nE|E(D_k^2|\cF_{k-1})-\sigma_k^2|/s_n^2&\rightarrow 0. \label{mclt.c2}
\end{eqnarray}

Since $E(Y_{ij}^6)<\infty$, by Proposition A.3 in \cite{portnoy1988asymptotic}, we have $E(\bY_k^T \bT_{k-1})^6\lesssim k^3p^3$  and therefore
$
\sum_{k=1}^n E|D_k|^3\lesssim n^{5/2}p^{3/2}+n^2p^{5/2}+n^4\norm{G}^3,
$

Together with Equation (\ref{s_n}) implies
$
\sum_{k=1}^nE|D_k|^3/s_n^3\rightarrow 0.
$

 Next, note that
 \begin{eqnarray}
 \sum_{k=1}^n E|E(D_k^2|\cF_{k-1})-\sigma_k^2|&\le&\left\{E\left[E(D_k^2|\cF_{k-1})-\sigma_k^2\]^2\right\}^{1/2} \\
 &\lesssim&n^{3/2}p^{3/2}+n^2p^{1/2}+n^{3/2}\norm{G}^2
 \end{eqnarray}
Thus, we have $\sum_{k=1}^n E|E(D_k^2|\cF_{i-1})-\sigma_k^2|/s_n^2\rightarrow 0$.
\end{proof}

\bigskip
Now we are able to prove the main results.

\begin{proof}[Proof of Theorem \ref{thm.noncen}] Under the alternatives, we have
\begin{eqnarray}
\Lambda_{\bgamma}&=&2n(\bthetahat_{\bgamma_f}-\bthetahat_{\bgamma})^T\overline{\bY}-2n\left[A(\bthetahat_{\bgamma_f})-A(\bthetahat_{\bgamma})\]\nonumber\\
&=&2n\left[(\bthetahat_{\bgamma_f}-\btheta^\ast)-(\bthetahat_{\bgamma}-\btheta_{\bgamma}^\ast)^T\]\overline{\bY}-2n\left[A(\bthetahat_{\bgamma_f})-A(\btheta_{\bgamma}^\ast)-A(\bthetahat_{\bgamma})+A(\btheta_{\bgamma}^\ast)\right]\nonumber\\
&~&-2n\left[(\btheta_{\bgamma}^\ast-\btheta^\ast)^T\overline{\bY}-A(\btheta_{\bgamma}^\ast)+A(\btheta^\ast)\].\label{3terms}
\end{eqnarray}
The first two terms of the LRT statistic in (\ref{3terms})  are approximately
$n\left(\norm{\overline{\bY}}^2-\norm{\overline{\bY}_{\bgamma}}^2\right)+O_p(p^2/n)$. Additionally, Lemma \ref{lem2} implies that the third term in (\ref{3terms}) is approximately equals to $-2n(\btheta_{\bgamma}^\ast-\btheta^\ast)\overline{\bY}+n\norm{\btheta_{\bgamma}^\ast-\btheta^\ast}^2$. Hence, applying Lemma \ref{lem3} completes the proof.
\end{proof}

\paragraph{Proof of Theorem \ref{thm3}.}

We show that if the sufficient condition in Theorem \ref{thm3} is satisfied, we have
$$
P\left(\bigcup_{\bgamma\in \Gamma_u}\{\Lambda_{\bgamma}\le q(\alpha; d_{\bgamma})\}\right)\rightarrow 0,\quad \text{as}\,\, n\rightarrow \infty.
$$

An union bound of the above probability is:
\begin{eqnarray}
P\left(\bigcup_{\bgamma\in \Gamma_u}\{\Lambda_{\bgamma}\le q(\alpha; d_{\bgamma})\}\right)&\le&\sum_{k=1}^{p-1}\sum_{\substack{d_{\bgamma}=k \\ \bgamma\in \Gamma_u}}P\big(\Lambda_{\bgamma}\le q(\alpha; k)\big)\\
&\le&\sum_{k=1}^{p-1} \exp\left[K_n(k)\right]\max_{\substack{d_{\bgamma}=k \\ \bgamma\in \Gamma_u}}P\left(\Lambda_{\bgamma}\le q(\alpha; k)\right).
\end{eqnarray}

Note that we assume the exponential bound as in (\ref{A0}) . Thus,  combining Lemma 8.1 in \citet{birge2001alternative} and Theorem A in \cite{inglot2010inequalities} gives the following probability upper bound:
\begin{eqnarray}\label{eq.upper.detect}
\max_{\substack{d_{\bgamma}=k \\ \bgamma\in \Gamma_u}}P\left(\Lambda_{\bgamma}(d_{\bgamma},\delta_{\bgamma})\le q(\alpha; k)\right)
&\le&\max_{\substack{d_{\bgamma}=k \\ \bgamma\in \Gamma_u}}P\left(X_{\delta_{\bgamma},k}\le q(\alpha,k)\right)+c_1\exp\left[-c_2 K_n(k)\right]\nonumber\\
&\le& \exp\left[-\min_{\substack{d_{\bgamma}=k\\\bgamma\in\Gamma_u}}\dfrac{\left(\delta_{\bgamma}+2\log(\alpha)-2\sqrt{-k\log(\alpha)}\right)^2}{2(k+2\delta_{\gamma})}\]\nonumber\\
&&\quad +c_1\exp\left[-c_2 K_n(k)\right]
\end{eqnarray}
where $c_1, c_2$ are positive constants. When $\min\limits_{\substack{d_{\bgamma}= k \\ \bgamma\in\Gamma_u}}\dfrac{\delta_{\bgamma}}{ K_n(k)}>B$ for some large enough positive constant $B$, the first term in the last upper bound (\ref{eq.upper.detect}) will be small than $\exp[-(c_3-1) K_n(k)]$ for some $c_3>1$.
Therefore, we have
\begin{eqnarray}
P\left(\bigcup_{\bgamma\in \Gamma_u}\{\Lambda_{\bgamma}\le q(\alpha; d_{\bgamma})\}\right)\lesssim \sum_{k=1}^{p-1}\exp[-\min(c_2,c_3-1) K_n(k)]\rightarrow 0.
\end{eqnarray}
\qed

\bigskip
\paragraph{Proof of Theorem \ref{thm.sharpness}} Let $\hat\bgamma$ be the largest model that are nested in all the models in MSCS. It may be the intercept only model. Note that with probability at least $1-\alpha$ the true model $\bgamma^\ast$ is included in the MSCS, i.e., $P\left(\bgamma^\ast\in \widehat{\Gamma}_\alpha\right)\ge 1-\alpha$. Therefore
\begin{eqnarray}
\sup_{\bbeta\in\cB}P_{\bbeta}\left(\hat\bgamma\neq\bgamma^\ast\right) &\le& \sup_{\bbeta\in\cB}P_{\bbeta}\left(\bgamma^\ast\notin \widehat{\Gamma}_\alpha\right)+\sup_{\bbeta\in\cB}P_{\bbeta}\left(\cD^{C}\right)\\
&\le& \alpha+1-\inf_{\bbeta\in\cB}P_{\bbeta}\left(\cD\right).
\end{eqnarray}
It follows that if $\inf_{\bbeta\in\cB}P_{\bbeta}\left(\cD\right)\rightarrow 1$ for any subsequence of $\{n_j\}\subset\{1,2,\cdots\}$, we must have $\sup_{\bbeta\in\cB}P_{\bbeta}\left(\hat\bgamma\neq\bgamma^\ast\right)\le \alpha'$ for some $\alpha< \alpha'< 1/2$ when $n_j$ is large enough in the subsequence. Hence if we can actually show  $\sup_{\bbeta\in\cB}P_{\bbeta}\left(\hat\bgamma\neq\bgamma^\ast\right)\ge 1/2$ for a small enough $c>0$ in the definition of $\cB$, the theorem is proved.

Note that this now becomes a traditional minimax framework where Fano's inequality can be applied. Without loss of generality, we assume $\sigma^2 = 1$ for the error variance. Consider a packing set $N_{\epsilon_n}$ in $\cB=\{\bbeta: \|\bbeta\|_0={r^\ast}\,\,\text{and}\,\,\|f_{\bbeta}\|_n^2\le cK_n({r^\ast})\}$ with packing distance $\epsilon_n$ being a small fraction of $\sqrt{cK_n({r^\ast})}$ under the $\|\|_n$ norm. Let $\bbeta$ be randomly chosen from the uniform distribution on $N_{\epsilon_n}$. Using similar arguments as in the proof of Theorem 11 in \cite{wang2014adaptive}, under the SRC, by choosing the constant $c$ small enough, the mutual information $\bI(\bbeta; \{X_i,Y_i\}_{i=1}^n)$ between the random ${\bbeta}$ and the observations
is upper bounded by $\delta=cK_n({r^\ast})/2$, and the local packing $\epsilon_n$-entropy $\log |N_{\epsilon_n}|$ is lower bounded by $2\delta+2\log2$. Apply Fano's inequality  \citep[see, e.g.,][]{yang1999information} to this linear regression model gives:
\begin{equation}
\sup_{\bbeta\in\cB}P_{\bbeta}\left(\hat\bgamma\neq\bgamma^\ast\right)\ge1-\dfrac{\bI(\bbeta; \{X_i,Y_i\}_{i=1}^n)+\log2}{\log |N_{\epsilon_n}|}\ge \dfrac{1}{2},
\end{equation}
this completes the proof.
\qed
\bigskip

\paragraph{Proof of Theorem \ref{coro1}} (i) Let $\bgamma$ be a model missing at least one variable in $\bgamma^\ast$.  From Theorem
\ref{thm3},
$P(\bgamma\in \widehat{\Gamma}_\alpha) \rightarrow 0$, as $n \rightarrow \infty$. Hence, for all $\theta_j$ in $\btheta^\ast$,
$\lim_{n\rightarrow\infty}P(II_j=1)=1$. This completes the first part of the theorem.

(ii) Let $\widetilde{\Gamma}=\{\bgamma_1,\bgamma_2,\cdots,\bgamma_N\}$ be
the set of models larger than $\bgamma^\ast$. The construction of MSCS implies $\lim_{n\rightarrow \infty}P(\bgamma_i\in \widehat{\Gamma}_\alpha)\ge 1-\alpha$  for any $1\le i\le N$.
Let $X_i$ be the random variable taking value $1$ when $\bgamma_i\in\widehat{\Gamma}_\alpha$ and $0$ otherwise. Note that when $k\notin\bgamma^\ast$ , there are $N/2$  models in $\widetilde{\Gamma}$ which contain $\theta_k$. Without loss of generality, let $\bgamma_1,
\bgamma_2,\cdots,\bgamma_{N/2}$ be models containing $k$. Let $Y_1=\sum_{i=1}^{N/2}X_i$,
denoting
the number of models in $\{\bgamma_1, \bgamma_2,\cdots,\bgamma_{N/2}\}$ that are included in $\widehat{\Gamma}_\alpha$. Similarly, let $Y_2=\sum_{i=N/2+1}^{N}X_i$.
Then $Y_1\le N/2$ and $E(Y_2)\ge N(1-\alpha)/2$. The $II_k$  defined in Section \ref{sec:RIW} equals to
$Y_1/(Y_1+Y_2)$ with probability going to $1$ as $n\rightarrow \infty$.
Note that for any $0<\Delta\le 1/2$ we have
\begin{align*}
P\left(\dfrac{Y_1}{Y_1+Y_2} \ge \dfrac{1}{2}+\Delta\right)&\le P\left(\dfrac{N/2}{N/2+Y_2}\ge\dfrac{1}{2}+\Delta\right)\\
&=P\left(N/2-Y_2\ge\dfrac{2\Delta N}{1+2\Delta}\right)\\
&\le \dfrac{(N/2-E(Y_2))(1+2\Delta)}{2N\Delta}\\
&\le \dfrac{\alpha(1+2\Delta)}{4\Delta},
\end{align*}
where the second inequality follows from Markov's inequality, which completes the proof.
\qed



\fontsize{11}{14pt}\selectfont
\renewcommand\bibname 

\clearpage







\end{document}